\documentclass[conference]{IEEEtran}
\IEEEoverridecommandlockouts
\usepackage{url}

\usepackage{breakurl}
\usepackage[linesnumbered,ruled,vlined]{algorithm2e}
\usepackage{cite}
\usepackage{subcaption}
\usepackage{amsmath,amssymb,amsfonts}
\usepackage{algorithmic}
\usepackage{romannum}
\usepackage{pdfpages}
\usepackage{graphicx}
\usepackage{psfrag}
\usepackage{textcomp}
\usepackage{xcolor}
\usepackage{geometry}
\usepackage{enumitem}
\usepackage{multicol}
\usepackage{lipsum} %
\usepackage{siunitx}
\usepackage{footmisc}
\usepackage{pgfplots}
\pgfplotsset{compat=newest}
\usetikzlibrary{plotmarks}
\usetikzlibrary{arrows.meta}
\usepgfplotslibrary{patchplots}
\usepackage{grffile}
\usepackage{amsmath}
\def\BibTeX{{\rm B\kern-.05em{\sc i\kern-.025em b}\kern-.08em T\kern-.1667em\lower.7ex\hbox{E}\kern-.125emX}}

\begin{document}
\title{Propagation Distance Estimation for Radio over Fiber with Cascaded Structure\\
}

\author{\IEEEauthorblockN{Dexin Kong,
  Diana Pamela Moya Osorio,
  and Erik G. Larsson}
\IEEEauthorblockA{\textit{Dept. of Electrical Engineering (ISY), Link\"oping University, Link\"oping, Sweden} \\}
\text{dexin.kong@liu.se, diana.moya.osorio@liu.se, erik.g.larsson@liu.se}
}
\maketitle
\begin{abstract}
Recent developments in polymer microwave fiber (PMF) have opened great opportunities for robust, low-cost, and high-speed sub-terahertz (THz) communications. Noticing this great potential, this paper addresses the problem of estimation of the propagation distance of a sub-Thz signal along a radio over fiber structure. Particularly, this paper considers a novel cascaded structure that interconnects multiple radio units (RUs) via fiber for applications in indoor scenarios. Herein, we consider the cascaded effects of distortions introduced by non-linear power amplifiers at the RUs, and the propagation channel over the fiber is based on measurements obtained from transmissions of sub-THz signals on high-density polyethylene fibers. For the estimation of the propagation distance, non-linear least-squares algorithms are proposed, and our simulation results demonstrate that the proposed estimators present a good performance on the propagation distance estimation even in the presence of the cascaded effect of non-linear PAs.

\end{abstract}
\vspace{0.2cm}

\section{Introduction}
The ever-increasing demand for wireless connectivity raises the necessity of pursuing the unexploited spectrum at very high frequencies. The opportunities to use the abundance of spectrum at the sub-terahertz (THz) frequency band have attracted attention for the sixth generation (6G)  wireless communication systems~\cite{THzsurvey,cai6G}. Although there has been abundant research on the performance of sub-THz wireless communication systems~\cite{THzdata,THzXR}, the implementation of sub-THz wireless communication systems is still an open challenge~\cite{THzimplementationchallenge}. Considering the cost of sub-THz radio hardware and the limited coverage~\cite{bjornson2019massive}, a low-cost and dense deployment solution is desirable.

\begin{figure}[t]
    \centering
    \includegraphics[width=1\linewidth]{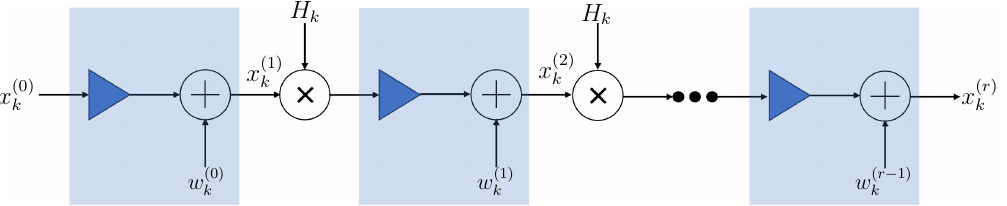}
    \caption{Signal model for the cascade ROF system}
    \label{signal model linear}
\end{figure}

Recent developments in polymer microwave fiber (PMF) have motivated the development of radio-over-fiber (ROF) communications, which offer great opportunities for low-cost implementations of sub-THz wireless  systems~\cite{analysisHE11,Foam-cladded,polymerblend}. Herein, we consider a  ROF system with a cascaded structure, operating at the sub-THz band, in which multiple components are interconnected via low-cost PMFs and finally connected to a central unit (CU). Fig.~\ref{signal model linear} illustrates the cascade ROF system under consideration (the different variables will be defined later). In this system, each user equipment (UE) transmits sub-THz signals over the air. And the signals are captured by densely deployed radio units (RUs) which are equipped with an antenna,  a power amplifier (PA), and other components. Then, the signals propagate over the ROF in a cascaded structure. Each RU in between the entry RU and the CU acts as a repeater which amplifies the signals and forwards them to the subsequent component, in a sequential process, until reaching the CU. All signal processing is then performed at the CU. 

For energy efficiency (EE) purposes, the PA is allowed to operate in the non-linear regime \cite{6Gwaveform}. Therefore, in each stage of the ROF system, the signals are distorted by the PA and by a segment of dispersive PMF. The fiber itself is a linear time-invariant system, whereas the amplifiers are non-linear. This characteristic enables us to estimate the propagation distance over the ROF by analyzing the way the signals are distorted when they propagate over the ROF. Our proposed propagation distance estimator could be a building block in a system for positioning the UE, for example, to help find the closest RU to the UE in the uplink (UL). 

\textbf{Contributions:} The specific contributions of this paper are 
(i) an analysis of the cascaded structure with non-linear PAs and dispersive PMFs, and the formulation of a corresponding UL signal model; (ii) a non-linear least-squares (NLS) optimization framework for propagation distance estimation, and
(iii) numerical results (using measured PMF dispersion characteristics and different values of non-linear factors) demonstrate that the proposed estimators present a good performance on the propagation distance estimation even in the presence of the cascaded effect of non-linear PAs.
To the best of our knowledge, signal propagation over a ROF with segments of dispersive PMFs and non-linear amplifiers connected in cascade has not been studied before. 
  
\begin{figure}[bt]
    \centering
    \includegraphics[width=0.9\linewidth]{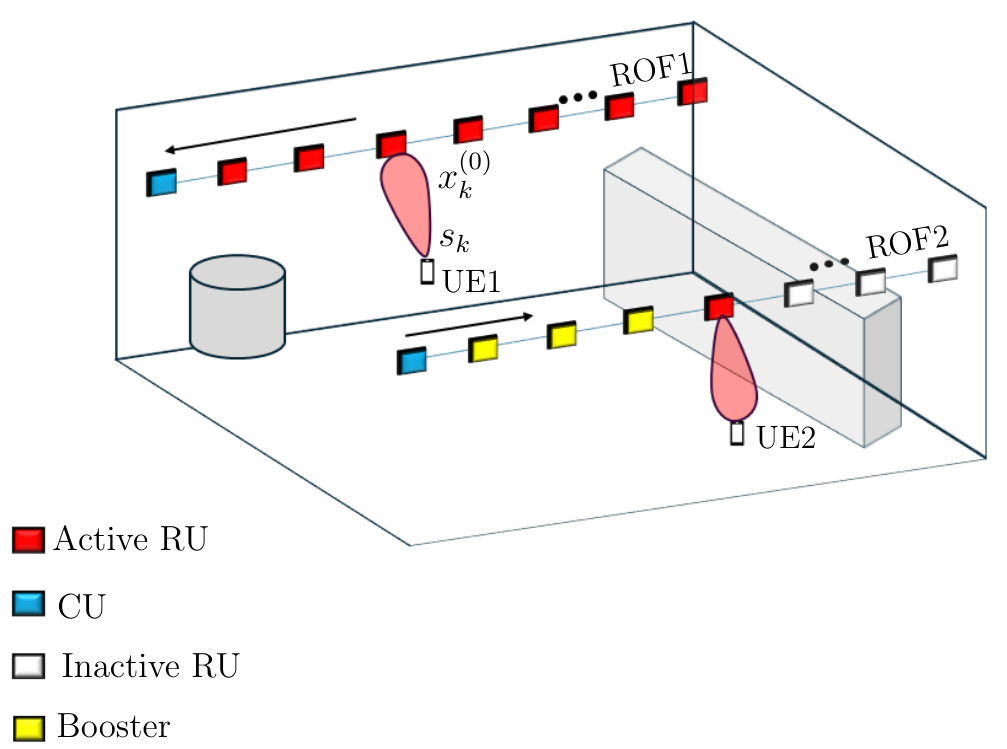}
    \caption{The deployment of ROF systems in an indoor scenario}
    \label{indoor scenario}
\end{figure}
   
\section{System Model and Problem Formulation}\label{PF and SM}

\subsection{System Model of a ROF System}
Consider an indoor scenario as illustrated in Fig.~\ref{indoor scenario}, where two UEs are served by two ROF systems, each with $M$ components interconnected by PMFs. The UE1 transmits data to ROF1 through the UL communication, where all RUs are active. The UE2 receives data from ROF2 through the DL communication, where the CU configures the RUs via the fiber according to the estimate of the UE2 position. Since the propagation range of the sub-THz wireless connection is extremely limited, we consider UE is only served by the nearest RU of the nearest ROF. 

\begin{figure}[bt]
    \centering\includegraphics[width=0.9\linewidth]{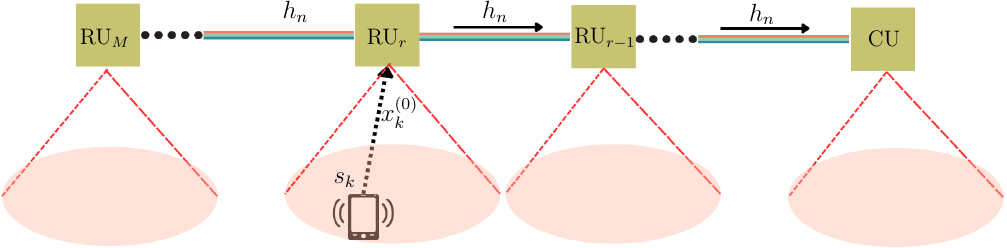}
    \caption{System model for UL communication between the UE1 and the ROF1}
    \label{radio stripe}
\end{figure}

It is considered that there is a clock offset and phase offset between the UEs and the ROF systems. Particularly, Fig.~\ref{radio stripe} illustrates the UL communication between UE1 and ROF1, where UE1 gets access through the unknown $r$th RU. The objective is to locate the nearest RU to the UE, i.e. estimating $r$, which leads to a good approximation of the UE location. In the following, we detail the two setups for the RUs. 

\subsection{RUs with PAs Operating in the Linear Regime}
Under this scenario, the RUs are equipped with PAs amplifying the input signals without any non-linearity. 
The signal model considers the channel over the fiber and the wireless channel from the UE to the RU. For a unit-length fiber segment, we model its impulse response via $L$ taps with amplitudes ($[\beta_0, \beta_1, \cdots, \beta_{L-1}]^\mathrm{T}$):
\begin{equation}
    h_{n}=\sum_{l=0}^{L-1} \beta_l \delta(n-l), \label{th}
\end{equation}
where $n$ is the time index and $\delta(\cdot)$ is the delta function. Let $D$ be the unknown propagation distance which is associated with an integer number $r$ indicating the number of unit lengths of the fiber that the signal has passed through, and can be expressed as $D=rd_f$,
where $d_f$ denotes the unit length of the fiber. Then, the impulse response can be expressed as  
\begin{equation}
    \Tilde{h}_{n}=\underbrace{h_{n} * h_{n}*\cdots*h_{n}}_{r\,\text{convolutions}},\label{hn}
\end{equation}
where all factors in \eqref{hn} are modeled as in~\eqref{th}. In the frequency domain, we consider the frequency response for a set of $K$ discrete frequencies, $\boldsymbol{f}=[f_0,\cdots,f_{K-1}]^\mathrm{T}$. The corresponding relation of \eqref{hn} in the frequency domain is in the form of multiplication of frequency responses. Let $\boldsymbol{H}=[H_0, \cdots, H_{K-1}]^\mathrm{T}$ represent the frequency response of a unit-length fiber. The frequency response of a unit-length fiber at the $k$th frequency $H_k$ is the Fourier transform of its impulse
\begin{equation}
        H_k= \sum_{n=0}^{N-1} h_n e^{-j2\pi f_k n T_s},
\end{equation}
where $T_s$ is the sampling time interval. Hence, the frequency response of an $r$-units-long fiber, at the $k$th frequency, can be expressed as 
\begin{equation}
    \Tilde{H}_k= \underbrace{H_k H_k \cdots H_k}_{r\,\text{multiplications}} = H_k^r.\label{H}
\end{equation} 

Considering~\eqref{hn} and~\eqref{H}, we aim at estimating $r$ of the propagation of a signal within the ROF. For that purpose, it is considered that the UE transmits a known sequence of samples $\boldsymbol{s}=[s_0, \cdots, s_{K-1}]^{\mathrm{T}} \in \mathbb{C}^{K\times1}$ over $K$ discrete frequencies. Initially, the signals undergo the wireless channel between the UE and the RU, which is unknown to the CU. 
Due to the sparse nature of the sub-THz wireless channel, only the line-of-sight components are considered in the signal model. The input signal for the $k$th frequency at the ROF, $x_k^{(0)}$ with $\boldsymbol{x}^{(0)}=[x_0^{(0)},\cdots, x_{K-1}^{(0)}]^{\mathrm{T}}$ (see Fig.~\ref{signal model linear}), can be written as 
\begin{equation}
    x_k^{(0)}=Ae^{-j2\pi f_k \tau}s_k, \label{xk0}
\end{equation}
where 
\begin{itemize}
    \item $A$ is a complex value absorbing the unknown wireless channel path gain, antenna gain, and the phase offset between the UE and the ROF.
    \item $\tau$ is the unknown propagation delay including the effects of one-way wireless signal propagation and the clock offset, which can be expressed as
    \begin{equation}
        \tau = \frac{d}{c} + \delta_{\text{t}}
    \end{equation}
    with $d$ being the distance over the wireless channel, $c$ is the speed of light, and $\delta_{\text{t}}$ is the clock offset between the UE and the ROF\footnote{Distinguish the effects of $d$ and $\delta_{\text{t}}$ is impossible because they have the same influence on the signals, i.e. phase shift and delay. However, one can rely on the effects of cascaded fibers and PAs to estimate $r$ and $\tau$ separately.}. 
\end{itemize}

The output signal at the $k$th frequency of the $r$th RU as $x_k^{(1)}$  is given as 
\begin{equation}
    x_k^{(1)}=G x_k^{(0)} + w_k^{(0)}.\label{xk1}
\end{equation}

In the rest of this paper, $G$ is the amplitude amplification of a PA, and $w^{(m)}_k \in \mathcal{CN}(0,\sigma^2)$ is the $k$th AWGN noise component at the $(r-m)$th RU. 


After propagating through $r$ RUs along the ROF system, the received signal with $\boldsymbol{x}^{(r)}=[x_0^{(r)},\cdots, x_{K-1}^{(r)}]^{\mathrm{T}}$ at the CU can be written as
\begin{equation}
    x_k^{(r)}=G^{r+1} H_k^{r} Ae^{-j2\pi f_k \tau} s_k + w_k^{(r-1)}.\label{xkr}
\end{equation}

The unknown parameters to be estimated in this problem are $A,r$, and $\tau$, of which the parameters $r$ and $\tau$ lead to a non-linear problem. We employ a non-linear least-squares (NLS) algorithm to tackle this estimation problem. First, we define the vector $\boldsymbol{g}\in \mathbb{C}^{K \times 1}$ as 
\begin{equation}
    \boldsymbol{g}=G^{r+1}[e^{-j2\pi f_0 \tau} H_0^{r} s_0,\cdots, e^{-j2\pi f_{K-1} \tau} H_{K-1}^{r} s_{K-1}]^\mathrm{T}. 
\end{equation}
The estimation objective can be written as:
\begin{equation}
\begin{bmatrix}
        \hat{r}\\
        \hat{\tau}\\
        \hat{A}
    \end{bmatrix}=\mathop{\min}_{r, \tau,A} \lvert\lvert \boldsymbol{x}^{(r)}- A \boldsymbol{g} \rvert\rvert^2,
\end{equation}
where $\lvert\lvert.\rvert \rvert$ denotes the $L^2$ norm. Dropping  irrelevant constants, the estimation of the nuisance parameter $A$ is a linear estimation problem that the estimate is obtained as 
\begin{equation}
\hat{A}= \frac{\boldsymbol{g}^{\mathrm{H}} \boldsymbol{x}^{(r)}}{\lvert\lvert\boldsymbol{g}\rvert\rvert^2}, \label{A_hat}
\end{equation}
where $(\cdot)^\mathrm{H}$ means the Hermitian transpose. After dropping the dependency on $A$ by considering $\hat{A}$, we can rewrite the objective as 
\begin{equation}\label{NLS}
    \begin{bmatrix}
        \hat{r}\\
        \hat{\tau}
    \end{bmatrix}=\mathop{\min}_{r, \tau} \bigg \lvert \bigg \lvert \boldsymbol{x}^{(r)}- \boldsymbol{g} (\boldsymbol{g}^{\mathrm{H}}\boldsymbol{g})^{-1}\boldsymbol{g}^{\mathrm{H}}\boldsymbol{x}^{(r)}\bigg \rvert \bigg \rvert^2. 
\end{equation}
Therefore, after a two-dimensional grid search, the estimate $\hat{r}$ as well as the nuisance parameter, $\hat{\tau}$, can be obtained. 
\subsection{RUs with PAs Operating in the Non-linear Regime}
In this section, we jointly consider the effects of non-linear PAs and fiber dispersion. 
The effects of non-linear amplifiers are characterized as a third-order polynomial by a factor $\lambda$ \cite{nonlinearPA,majidi2014analysis}. Again, the UE transmits a known sequence of samples $\boldsymbol{s}=[s_0,\cdots,s_{K-1}]^\mathrm{T}$ over $K$ discrete frequencies to the ROF system. We model the input signal in the time domain as

\begin{equation}
    x_n^{(0)}=\frac{1}{K}\sum_{k=0}^{K-1} e^{j2\pi\frac{n}{N} k} A e^{-j 2\pi f_k \tau}s_k,\label{xn0}
\end{equation}
with $x_n^{(0)} \in [x_0^{(0)},\cdots, x_{K-1}^{(0)}]^\mathrm{T}.$

The output of the entry RU can be written as  
\begin{equation}
x_{n}^{(1)}=G\bigg(x_n^{(0)}+\!\!\underbrace{\lambda x_n^{(0)} \left\lvert x_n^{(0)} \right \rvert^2}_{\text{PA's non-linearity}}\!\!\!\bigg)+w_{n}^{(0)}, \label{xn1}
\end{equation}
where $w^{(0)}_n \in \mathcal{CN}(0,\sigma^2)$ is the $n$th AWGN noise component at the $r$th RU. To express the output at the following RUs, we define the function $f(\cdot)$ that combines the effects of the channel dispersion and the non-linearity and can be written as
\begin{equation}
        f(x_{n}^{(1)}) = G\left(\sum_{l=0}^{L-1} \beta_l x_{n-l}^{(1)}+\lambda \sum_{l=1}^{L-1} \beta_l x_{n-l}^{(1)} \left\lvert \sum_{l=1}^{L-1} \beta_l x_{n-l}^{(1)} \right\rvert^2\right). \label{ffunction} 
\end{equation}
If $f(\cdot)$ takes a vector as input, it performs \eqref{ffunction} element-wise. 

The signal $\boldsymbol{x}^{(1)}$ would undergo $(r-1)$ RUs to reach the CU. 
Consequently, the signal $\boldsymbol{x}^{(1)}$ would undergo the recursive process $r$ times. Let $f^{r}(\boldsymbol{x}^{(1)})$ be the overall function, which can be expressed as $f^{r}(\boldsymbol{x}^{(1)})=\underbrace{f(f(f(...(\boldsymbol{x}^{(1)}))))}_{r \, \text{times}}$.


Therefore, the received signal at the CU can be expressed as $\boldsymbol{x}^{(r)}= f^{r}\left(\boldsymbol{x}^{(1)}\right)+\boldsymbol{w}^{(r-1)}$. Following the LS framework, the objective function w.r.t. the unknown parameter $A,r,$ and $\tau$ is given by 
    \begin{equation}
        \begin{bmatrix}
        \hat{A}\\
        \hat{\tau}\\
        \hat{r}
    \end{bmatrix}=\mathop{\min}_{A,\tau,r} \bigg\lvert\bigg\lvert\boldsymbol{x}^{(r)}-f^{r}\left(G\left(x_n^{(0)}+\lambda x_n^{(0)} \left\lvert x_n^{(0)} \right \rvert^2\right)\right)\bigg\rvert\bigg\rvert^2.\label{ml}
\end{equation}
To reduce the complexity of the three-dimensional grid search, we propose Algorithm~\ref{Grid search}, which is a coordinate descent algorithm, to solve this problem.  




\section{Simulations and Results}\label{SR}
\begin{figure}[tbp]
    \centering
    \includegraphics[width=0.8\linewidth]{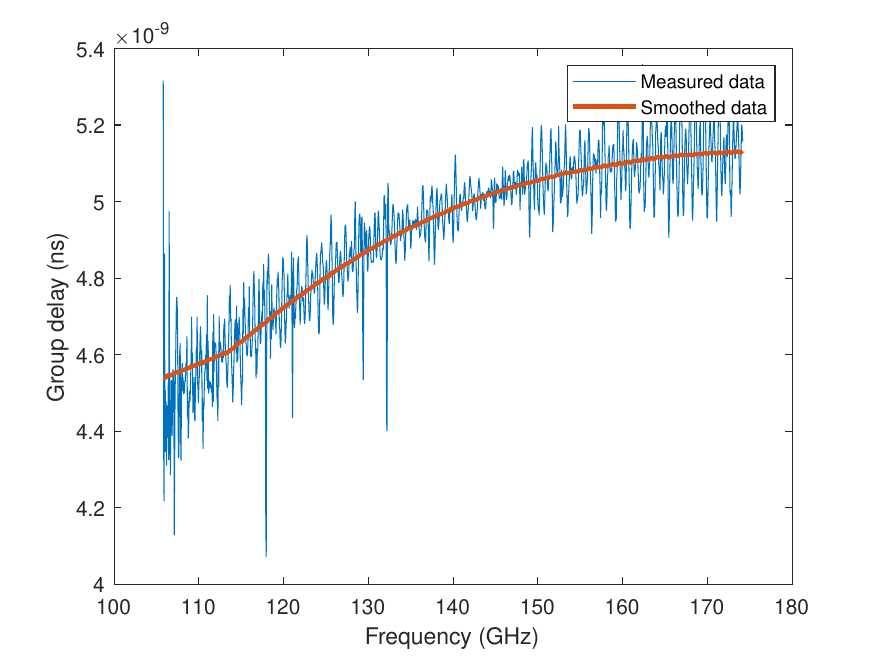}
    \caption{Group delay}
    \label{gd}
\end{figure}

\begin{figure}[tbp]
    \centering
    \includegraphics[width=0.8\linewidth]{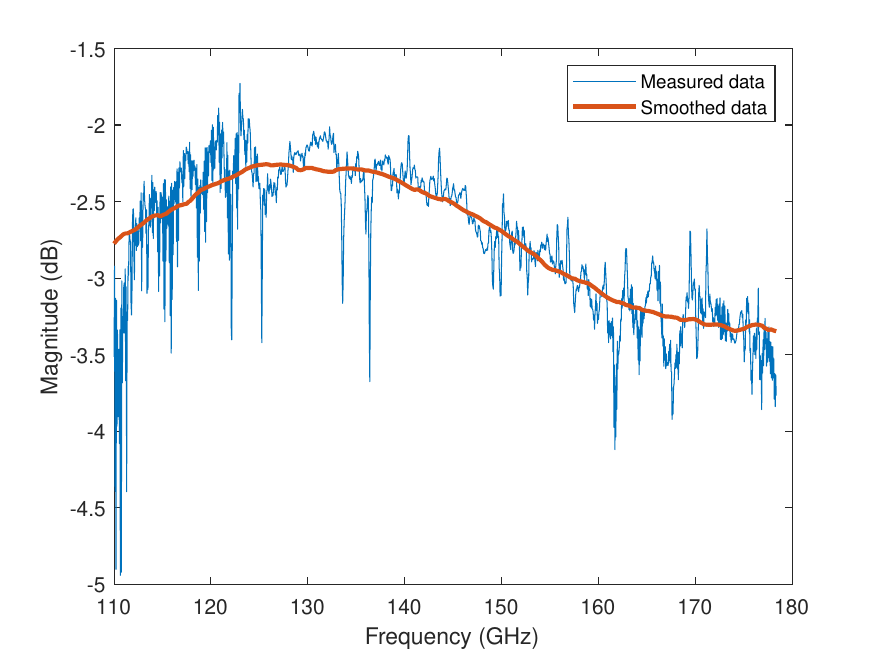}
    \caption{Magnitude}
    \label{pl}
\end{figure}

\begin{figure}[tbp]
    \centering
    \includegraphics[width=0.8\linewidth]{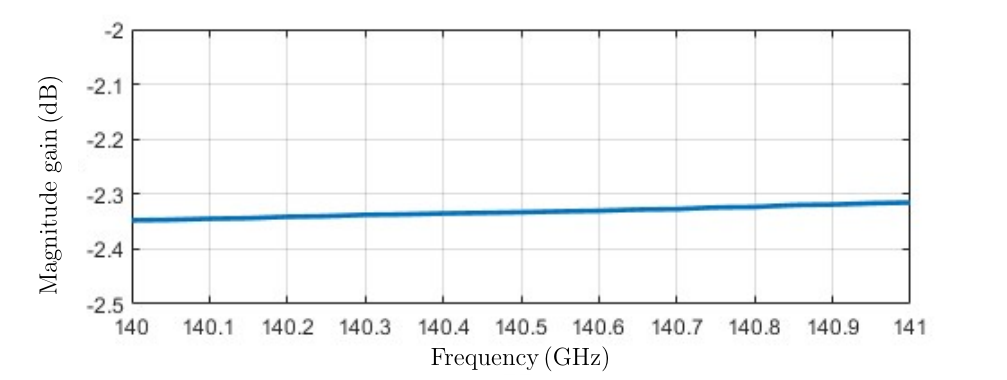}
    \caption{Channel transfer function}
    \label{channeltf}
\end{figure}

In this section, we illustrate the performance obtained by the proposed estimators. We use the measurements presented in~\cite{Frida}, for a 1-meter PMF made out of high-density polyethylene with a solid rectangular cross-section in the D-band (110 GHz to 170 GHz). The measured data consists of channel characteristics at $K$ discrete frequencies, encompassing both magnitudes ($\boldsymbol{\alpha}=\big[\lvert H_0\rvert, \cdots, \lvert H_{K-1} \rvert\big]^\mathrm{T}$) and group delay. These measurements inherently include several imperfections, such as noise, interference, and measurement errors. Therefore, the Matlab built-in \textit{smoothdata} function was employed to average out these impairments as shown in Figs.~\ref{gd} and~\ref{pl}. Subsequently, the phase response $\boldsymbol{\phi}=[\phi_0, \cdots, \phi_{K-1}]^\mathrm{T}$ was obtained by integrating the group delay on the discrete frequencies. Then, the transfer function can be expressed as
\begin{equation}
    \boldsymbol{H}= \boldsymbol{\alpha} \circ \exp{(j \boldsymbol{\phi)}},\label{MeasuredH}
\end{equation}
where $\circ$ represents the Hadamard product. This channel corresponds to a spectrum with 1 GHz bandwidth centered around 140 GHz, as illustrated in Fig.~\ref{channeltf}. 
\begin{figure}[tbp]
    \centering
    \includegraphics[width=0.9\linewidth]{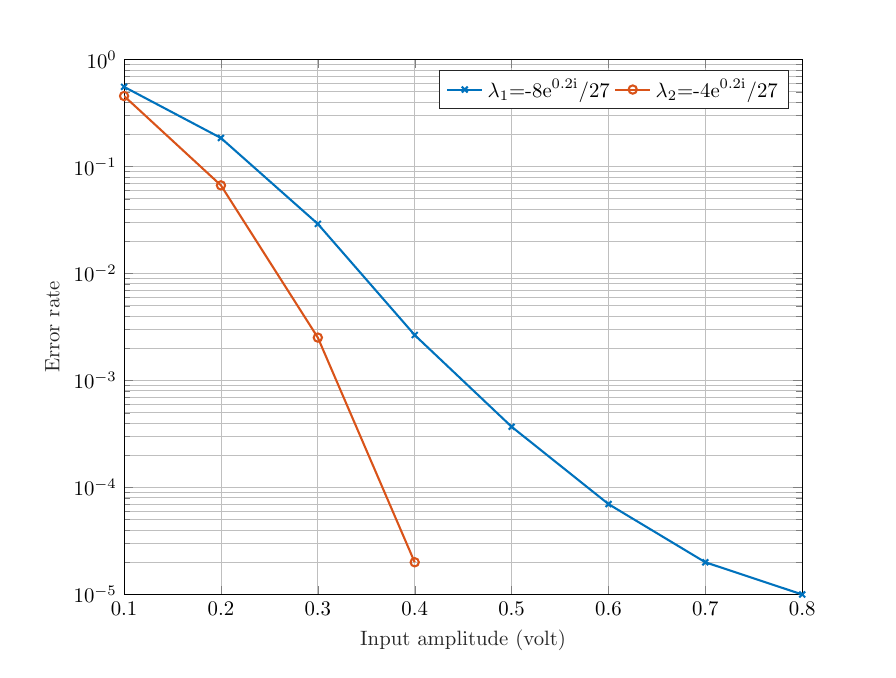}
    \caption{Error rate of $\hat{r} \neq r$ in a ROF system}
    \label{r_non-linear}
\end{figure}

With PAs operating in both linear and non-linear regimes, we simulate a ROF system comprising five RUs and one CU. Between every two components, a PMF with the frequency response as expressed in \eqref{MeasuredH} serves as the connection. The UE transmits a known sequence of samples using a quadrature phase shift keying (QPSK) waveform. The UE accesses the ROF system through the third RU. Monte Carlo simulations were performed to demonstrate the performance of the estimators in \eqref{NLS} and \eqref{ml}. In this paper, 
the noise at each stage is white over the whole bandwidth with variance $\sigma^2$ is 0\,dB. 

From Fig.~\ref{channeltf}, the maximum magnitude attenuation in a one-meter fiber is approximately 2.4\,dB. Therefore, the amplification factor $G$ is set to 2.4\,dB. Monte Carlo simulations were also performed with the distortion incurred by PAs operating with two non-linear factors, i.e. $\lambda_1= -\frac{4}{27}e^{j0.2},$ and $\lambda_2= -\frac{8}{27}e^{j0.2}$. 
The proposed grid search algorithm was implemented in Matlab and the results are shown in Fig.~\ref{r_non-linear}. 
The configurations of the antenna gain and the path loss in the wireless channel were set to ensure the input amplitude is in the certain regime (linear or non-linear) of the PAs. It is observed that the error rate decreases with increased input amplitude, which indicates that the proposed algorithm is able to accurately estimate the propagation distance even when PAs work in the non-linear regime. Through comparison between the estimation accuracy with two non-linear factors, one can observe an accuracy loss with a higher non-linear factor.  

\section{Conclusion}
We considered a novel, low-cost, and easily deployable sub-THz communication system implementation that combines PMFs and PAs in a cascaded structure. An UL signal model of this system was developed, both with and without  PA non-linearities. Based on this model, a NLS framework was developed for the estimation
of propagation distance along the ROF of signals transmitted from a UE. The NLS objective was minimized by using a combination of coordinate descent and a grid search. Monte-Carlo simulations were performed to assess the performance of the proposed estimator. 
Results show that a good performance of the propagation distance estimation can be attained under the effects of cascaded dispersive fibers and non-linear PAs. 
Our proposed algorithms show the feasibility of localizing UEs in the proposed system and provide a starting point for the development of high-resolution positioning schemes.

Future work can be conducted by considering the presence of automatic gain control loop, which makes the gain of PAs adaptive. 
\begin{algorithm}[hpbt]
 \KwData{$\boldsymbol{s}, f(x)$, $\boldsymbol{x}^{(r)}$, $T$}
 \KwResult{$\hat{r}, \hat{\tau}, \hat{A}$}
 initialization:$\lvert A \rvert \in (0,1]$, $\angle A \in[-\pi,\pi]$, $\tau \in 1:1:10$\,ns, $r \in [1,2,3,4,5]$,  $cost=c$\, $\omega= \bigg\lvert\bigg\lvert\boldsymbol{x}^{(r)}- f^{r}\left(\boldsymbol{x}^{(1)}\right) \bigg\rvert \bigg \rvert^2$\; 
 \While{$cost \leq T$}{
     \For{$\ \tau  \gets 1$ \KwTo $10$\,ns}{
    \eqref{ml}\;
    \eqref{xn1}\;
     $cost_1(\tau)=\omega$;
   end}
   $\tau=\mathop{\min}_{\tau}cost_1$\;
   \For{$\lvert A \rvert \gets 0.1$ \KwTo $1$}{
   \eqref{xn0}\; 
    \eqref{xn1}\;
     $cost_2(\lvert A \rvert)=\omega$\;
   end}
   $\lvert A \rvert=\mathop{\min}_{\lvert A \rvert}cost_2$\;
   \For{$\ \angle A \gets -\pi$ \KwTo $\pi$}{
   \eqref{xn0}\;
    \eqref{xn1}\;
     $cost_3( \angle(A))=\omega$\;
   end}
   $\angle A =\mathop{\min}_{\angle A }cost_3$\;
 \For{$r \gets 1$ \KwTo $5$}{
   \eqref{xn0}\;
   \eqref{xn1}\;
    $cost_4(r)=\omega$\;
   end}
   $r=\mathop{\min}_{r}cost_4$\;
   \eqref{xn0}\;
   \eqref{xn1}\;
     $cost=\omega$\;
   }
  end
 \caption{Grid search}
 \label{Grid search}
\end{algorithm}
\section*{Acknowledgment}
This work was supported in part by ELLIIT, and in part by the 6G Tandem project funded by European Union’s Horizon Europe research and innovation programme under Grant Agreement No 101096302.



\bibliographystyle{IEEEtran}

\begin{thebibliography}{10}
\providecommand{\url}[1]{#1}
\csname url@samestyle\endcsname
\providecommand{\newblock}{\relax}
\providecommand{\bibinfo}[2]{#2}
\providecommand{\BIBentrySTDinterwordspacing}{\spaceskip=0pt\relax}
\providecommand{\BIBentryALTinterwordstretchfactor}{4}
\providecommand{\BIBentryALTinterwordspacing}{\spaceskip=\fontdimen2\font plus
\BIBentryALTinterwordstretchfactor\fontdimen3\font minus \fontdimen4\font\relax}
\providecommand{\BIBforeignlanguage}[2]{{%
\expandafter\ifx\csname l@#1\endcsname\relax
\typeout{** WARNING: IEEEtran.bst: No hyphenation pattern has been}%
\typeout{** loaded for the language `#1'. Using the pattern for}%
\typeout{** the default language instead.}%
\else
\language=\csname l@#1\endcsname
\fi
#2}}
\providecommand{\BIBdecl}{\relax}
\BIBdecl

\bibitem{THzsurvey}
Z.~Chen, X.~Ma, B.~Zhang, Y.~Zhang, Z.~Niu, N.~Kuang, W.~Chen, L.~Li, and S.~Li, ``A survey on {Terahertz} communications,'' \emph{China Communications}, vol.~16, no.~2, pp. 1--35, 2019.

\bibitem{cai6G}
X.~Cai, X.~Cheng, and F.~Tufvesson, ``Toward 6{G} with terahertz communications: Understanding the propagation channels,'' \emph{IEEE Communications Magazine}, vol.~62, no.~2, pp. 32--38, 2024.

\bibitem{THzdata}
H.~Elayan, O.~Amin, R.~M. Shubair, and M.-S. Alouini, ``{Terahertz} communication: The opportunities of wireless technology beyond 5{G},'' in \emph{2018 International Conference on Advanced Communication Technologies and Networking (CommNet)}, 2018, pp. 1--5.

\bibitem{THzXR}
C.~Chaccour, W.~Saad, O.~Semiari, M.~Bennis, and P.~Popovski, ``Joint sensing and communication for situational awareness in wireless {THz} systems,'' \emph{arXiv preprint arXiv:2111.14044}, 2021.

\bibitem{THzimplementationchallenge}
P.~Heydari, ``{Terahertz} integrated circuits and systems for high-speed wireless communications: Challenges and design perspectives,'' \emph{IEEE Open Journal of the Solid-State Circuits Society}, vol.~1, pp. 18--36, 2021.

\bibitem{bjornson2019massive}
E.~Bj{\"o}rnson, L.~Van~der Perre, S.~Buzzi, and E.~G. Larsson, ``{Massive} {MIMO} in sub-6 {GHz} and {mmWave}: Physical, practical, and use-case differences,'' \emph{IEEE Wireless Communications}, vol.~26, no.~2, pp. 100--108, 2019.

\bibitem{analysisHE11}
A.~Standaert, M.~Rousstia, S.~Sinaga, and P.~Reynaert, ``Analysis and experimental verification of the {HE11} mode in hollow {PTFE} fibers,'' in \emph{2015 Asia-Pacific Microwave Conference (APMC)}, vol.~3.\hskip 1em plus 0.5em minus 0.4em\relax IEEE, 2015, pp. 1--3.

\bibitem{Foam-cladded}
M.~De~Wit, Y.~Zhang, and P.~Reynaert, ``Analysis and design of a foam-cladded {PMF} link with phase tuning in 28-nm {CMOS},'' \emph{IEEE Journal of solid-state circuits}, vol.~54, no.~7, pp. 1960--1969, 2019.

\bibitem{polymerblend}
P.~Reynaert, M.~Tytgat, W.~Volkaerts, A.~Standaert, Y.~Zhang, M.~De~Wit, and N.~Van~Thienen, ``Polymer microwave fibers: a blend of {RF}, copper and optical communication,'' in \emph{ESSCIRC Conference 2016: 42nd European Solid-State Circuits Conference}.\hskip 1em plus 0.5em minus 0.4em\relax IEEE, 2016, pp. 15--20.

\bibitem{6Gwaveform}
M.~Sarajlić, N.~Tervo, A.~Pärssinen, L.~H. Nguyen, H.~Halbauer, K.~Roth, V.~Kumar, T.~Svensson, A.~Nimr, S.~Zeitz, M.~Dörpinghaus, and G.~Fettweis, ``Waveforms for sub-{THz} 6{G}: Design guidelines,'' in \emph{2023 Joint European Conference on Networks and Communications \& 6G Summit (EuCNC/6G Summit)}, 2023, pp. 168--173.

\bibitem{nonlinearPA}
S.~R. Aghdam, S.~Jacobsson, U.~Gustavsson, G.~Durisi, C.~Studer, and T.~Eriksson, ``Distortion-aware linear precoding for massive {MIMO} downlink systems with nonlinear power amplifiers,'' \emph{arXiv preprint arXiv:2012.13337}, 2020.

\bibitem{majidi2014analysis}
M.~Majidi, A.~Mohammadi, and A.~Abdipour, ``Analysis of the power amplifier nonlinearity on the power allocation in cognitive radio networks,'' \emph{IEEE Transactions on Communications}, vol.~62, no.~2, pp. 467--477, 2014.

\bibitem{Frida}
\BIBentryALTinterwordspacing
F.~Strömbeck, ``\BIBforeignlanguage{{eng}}{{Integrated Circuit Design for High Data Rate Polymer Microwave Fiber Communication}},'' {2023}, {Doctoral thesis}. [Online]. Available: \url{{https://research.chalmers.se/en/publication/534754}}
\BIBentrySTDinterwordspacing

\end{thebibliography}
\end{document}